\documentclass[twocolumn,showpacs,preprintnumbers]{revtex4}
\usepackage{amssymb}
\usepackage{amsmath}
\usepackage{graphicx}
\usepackage{dcolumn}
\usepackage{bm}

\begin{document}

\preprint{APS/123-QED}
\title{Decoherence-free quantum memory for photonic state using atomic
ensembles\\
}
\author{Feng Mei}
\author{Ya-Fei Yu}
\email{yfyuks@hotmail.com}
\author{Zhi-Ming Zhang}
\email{zmzhang@scnu.edu.cn}
\affiliation{Laboratory of Photonic Information Technology, School for Information and
Photoelectronic Science and Engineering, South China Normal University,
Guangzhou 510006, PR China}
\date{November 4, 2008}

\begin{abstract}
Large scale quantum information processing requires stable and long-lived
quantum memories. Here, using atom-photon entanglement, we propose an
experimentally feasible scheme to realize decoherence-free quantum memory
with atomic ensembles, and show one of its applications, remote transfer of
unknown quantum state, based on laser manipulation of atomic ensembles,
photonic state operation through optical elements, and single-photon
detection with moderate efficiency. The scheme, with inherent
fault-tolerance to the practical noise and imperfections, allows one to
retrieve the information in the memory for further quantum information
processing within the reach of current technology.
\end{abstract}

\maketitle

% Force line breaks with \\



Quantum information science involves the storage, manipulation and
communication of information encoded in quantum systems. In the future, an
outstanding goal in quantum information science is the faithful mapping of
quantum information between a stable quantum memory and a reliable quantum
communication channel. The quantum memory is a key element of quantum
repeaters \cite{1} that allow for long-distance quantum communication over
realistic noisy quantum channel, which is also necessary for scalable linear
optics quantum computation put forward by Knill et al. \cite{2}. Atomic
systems are excellent quantum memories, because appropriate internal
electronic states can coherently store qubits over very long timescales.
Photons, on the other hand, are the natural platform for the distribution of
quantum information between remote qubits, by considering their ability to
transmit large distance with little perturbation \cite{3}. Recently, various
quantum memory schemes for storing photonic quantum states have been
proposed, involving employing one atom or two atoms in a high-Q cavity \cite%
{4}, all optical approaches \cite{5}, and coupling the light into the atomic
ensembles \cite{6}. However, the above schemes have several disadvantages.
It is hard to efficiently couple a photon with a atom in a high-fineness
cavity, all optical approaches have large transmission loss. Given these
drawbacks it is of interest to explore the alternatives.

In this paper, different from the quantum memory schemes using
electromagnetically induced transparency (EIT) \cite{6}, we propose a scheme
to achieve the decoherent-free quantum memory with atom ensembles and linear
optics by quantum teleportation. By using the photonic qubits as the
information carriers and the collective atomic qubits as the quantum memory,
the information can be encoded into the remote decoherence-free subspaces
(DFS) \cite{7,8} of quantum memory via teleportation of an arbitrarily
prepared quantum state. Compared with the above proposals, our protocol has
the following significant advances: (i) It is not necessary to employ a
cavity which works in the strong coupling regime. Instead, sufficiently
strong interaction is achieved due to the collective enhancement of the
signal-to-noise ratio \cite{9,10,11}. (ii) Different from all optical
approaches, the fidelity of our atomic memory is insensitive to photon loss.
The photon loss only influences the probability of success. (iii) We use two
atomic ensembles to encode a single logic qubit, the information is stored
in a decoherent-free subspace (DFS) which can protect quantum information
against decoherence effectively. (iv) Raman transition technique provides a
controllable and decoherence-insensitive way of coupling between light and
atoms, so the information storing in the quantum memory can be retrieved
with ease for further quantum information processing. (v) Finally, by
introducing the photon an additional degree of freedom with spatial modes,
we get complete Bell state measurement (BSM) and improve the preparation
efficiency a lot.

Before describing the detailed model, first we summarize the basic ideas of
the scheme, which consists of four steps shown in Fig. 1. (A) Firstly, the
entanglement between atomic ensembles and Stokes photon is generated. (B)
The spatial mode is employed as additional degree of freedom of photon to
encode the information we want to imprint. (C) Next, complete Bell state
measurement is performed on the polarization and spatial qubits of photon.
(D) Via the BSM results, the information will be faithfully encoded into the
decoherent-free subspace (DFS) of atomic ensembles.

\textit{Step }(A) --- Entanglement between atom ensembles and photon \cite%
{12} is crucial to achieve this task. In more detail, the basic element is a
cloud of $N$ identical atoms with the relevant level structure shown in Fig.
1(a). The metastable lower states $\left\vert g\right\rangle $ and $%
\left\vert s\right\rangle $ can correspond to hyperfine or Zeeman sublevels
of the ground state of alkali-metal atoms, which have an experimentally
demonstrated long lifetimes \cite{13,14}. To achieve effectively enhanced
coupling to light, the atom ensembles should be preferably placed with
pencil-shape. Initially, all the atoms are prepared in the ground state $%
\left\vert g\right\rangle $. Shining a synchronized short, off-resonant pump
pulse\emph{\ }into the atomic ensemble $j\left( j=L\text{ or }R\right) $
induces Raman transitions into the state $\left\vert s\right\rangle $. The
emission of single Stokes photon results in the state $S_{j}^{+}\left\vert
0_{a}\right\rangle _{j}$ of atomic ensembles, where the ensemble ground
state $\left\vert 0_{a}\right\rangle $ $=\otimes _{i}\left\vert
g\right\rangle _{i},$the symmetric collective mode $S=$ $\left( 1/\sqrt{N_{a}%
}\right) \sum_{i}\left\vert g\right\rangle _{i}\left\langle s\right\vert $
\cite{15}. By the selection rules and the conservation of angular momentum,
the pump pulse and the Stokes photon have the left $\left( L\right) $ and
right $\left( R\right) $ circular polarization. Assume that the interaction
time is short, so, the mean number of the forward-scattered Stokes photon is
much smaller than 1. We can define signal light mode bosonic operator $a$
for the Stokes pulse with its vacuum state denoted by $\left\vert
0_{p}\right\rangle $, where $a^{+}\left\vert 0_{p}\right\rangle =\left\vert
R\right\rangle $. The symmetric collective mode $S$ and the signal light
mode $a$ are correlated with each other, which means, if the atomic ensemble
is excited to the symmetric collective mode $S^{+}$, the accompanying
emission photon will go to the signal light mode $a^{+}$, and vice versa.
The whole state of atomic ensemble and the Stokes photon can be written as \
\ \ \ \ \ \ \ \ \ \ \ \ \ \ \ \ \ \ \ \ \ \ \ \ \
\begin{equation}
\left\vert \phi \right\rangle _{j}=\left\vert 0_{a}0_{p}\right\rangle _{j}+%
\sqrt{p_{cj}}S_{j}^{+}a_{j}^{+}\left\vert 0_{a}0_{p}\right\rangle
_{j}+o(p_{cj}),  \label{1}
\end{equation}%
where $p_{cj}=4g_{c}^{2}N_{j}L_{j}/c\left\vert \Omega \right\vert ^{2}/%
\mathbf{\bigtriangleup }^{2}t_{p}$ is the small excitation probability of
single spin flip in the ensemble $j$ \cite{15,16}. Here $g_{c}$ is
atom-field coupling constant, $N_{j}$ and $L_{j}$ are the linear density and
the length of the atomic ensemble $j$. The probability can be controlled by
adjusting the light-atom interaction time and pulse duration $t_{p}$. $%
o(p_{cj})$ represents its more excitations whose probabilities are equal to
or smaller than $p_{cj}^{2}$ \cite{15}. We should note that the scattered
photon goes to some other optical modes other the signal mode. However, when
$N$ is large, the independent spontaneous emissions distribute over all the
atomic modes, whereas the contribution to the signal light mode will be
small \cite{10}. So the use of atomic ensembles will result in a large
signal-to-noise ratio \cite{9,10,11} and improve the efficiency of the
scheme.

\begin{figure}[tbp]
\includegraphics [width=9cm,height=4cm]{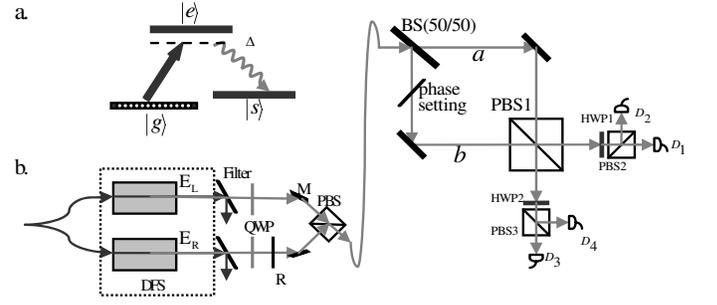}
\caption{Schematic setup for quantum memory scheme. (a) The relevant atomic
level structure in the ensembles with $\left\vert e\right\rangle $ the
excited state, $\left\vert g\right\rangle $ the ground state and $\left\vert
s\right\rangle $ the metastable state for storing a qubit of information.
The transition $\left\vert g\right\rangle \rightarrow \left\vert
e\right\rangle $ can be coupled through the left circular classical pump
pulse with the Rabi frequency\textit{\ }$\Omega $ and a detuning $\mathbf{%
\bigtriangleup }$, and the forward-scattered Stokes light comes from the
transition $\left\vert e\right\rangle \rightarrow \left\vert s\right\rangle $%
, which has a right circular polarization. (b) The forward-scattered Stokes
fields are collected after the filters which are polarization- and
frequency- selective to filter the pumping light, and interfere at PBS. The
output Stokes photon is coupled into a single-mode optical fiber and guided
to the setup for preparing the information and subsequent
polarization-spatial Bell-state measurement. The phase setting $\left(
\protect\alpha ,\protect\beta \right) $ allows to prepare any superposition
of the spatial-mode qubit. Based on the PBSs and single-photon detectors, we
can get the BSM in the polarization-spatial-mode Hilbert space of the Stokes
photon.}
\end{figure}

We can make $p_{cL}=p_{cR}$, the two emitted Stokes pulses are interfered on
a polarized beam splitter (PBS) after transmitting a quarter wave plate
(QWP) shown in Fig. 1(b). The PBS transmits only H and reflects V
polarization component, the QWP transforms the circularly polarized photon
into linearly polarized photon by the operator $P_{H}^{+}=\left\vert
H\right\rangle \left\langle R\right\vert $ and $P_{V}^{+}=\left\vert
V\right\rangle \left\langle L\right\vert $. The small fraction of the
transmitted classical pulses can be easily filtered through the filters. For
the Stokes pulse from the atomic ensemble $R$, a polarization rotator
\textbf{R} is inserted after the QWP. The function of the rotator is defined
by the operator $\overset{\sim }{R}=\left\vert H\right\rangle \left\langle
V\right\vert +$ $\left\vert V\right\rangle \left\langle H\right\vert $. By
selecting orthogonal polarization, conditional on a single-photon detector
click \cite{17}, the whole state of the atomic ensembles and the Stokes
photon evolves into a maximal entangled state

\begin{equation}
\left\vert \Psi \right\rangle _{ap}=\left( S_{L}^{+}+S_{R}^{+}\overset{\sim }%
{R}\right) /\sqrt{2}\left\vert vac\right\rangle _{ap}\text{.}  \label{2}
\end{equation}%
Because during the Stokes photons from the ensemble $L$ and $R$ interfere in
the input of the PBS, the information of their paths is erased. It denotes $%
\left\vert vac\right\rangle _{ap}\equiv \left\vert 0_{a}\right\rangle
_{L}\left\vert 0_{a}\right\rangle _{R}\left\vert H\right\rangle $. Then, the
entangled state can be rewritten as
\begin{equation}
\left\vert \Psi \right\rangle _{ap}=\left( \left\vert H\right\rangle
\left\vert 1\right\rangle _{a}+\left\vert V\right\rangle \left\vert
0\right\rangle _{a}\right) /\sqrt{2}\text{,}  \label{3}
\end{equation}%
where $\left\vert 0\right\rangle _{a}=\left\vert 0_{a}\right\rangle
_{L}\left\vert 1_{a}\right\rangle _{R}$, $\left\vert 1\right\rangle
_{a}=\left\vert 1_{a}\right\rangle _{L}\left\vert 0_{a}\right\rangle _{R}$
denote one spin flip in one of the ensembles. In fact, the generated
entangled state will be mixed with a small vacuum component if we take into
account the detector dark counts. However, the vacuum component is typically
much smaller than the repetition frequency of the Raman pulses. Here, we can
neglect this small vacuum component and the high order terms.

In the following, we will use the atomic ensembles qubit of the atom-photon
entanglement as our quantum memory. Long-lived quantum memory is the
keystone of quantum repeater. Unfortunately, due to environmental coupling,
the stored information can be destroyed, so-called decoherence. Here,
decoherence-free subspace (DFS) has been introduced to protect fragile
quantum information against detrimental effects of decoherence. To establish
the DFS, we utilize the state of a pair of atomic ensembles to encode a
single logic qubit, i.e., $\left\vert 0\right\rangle _{a}=\left\vert
0_{a}\right\rangle _{L}\left\vert 1_{a}\right\rangle _{R}$, $\left\vert
1\right\rangle _{a}=\left\vert 1_{a}\right\rangle _{L}\left\vert
0_{a}\right\rangle _{R}$, so that the phase noise can be effectively
suppressed.

\textit{Step }(B) --- After the generation of\ the entanglement, the emitted
Stokes photon is coupled into a fiber and guided to the setup illustrated in
Fig. 1(b), where the state we want to imprint into the quantum memory is
prepared. The Hilbert space of the photon is extended by using two spatial
modes as an additional degree of freedom \cite{18,19}. The photon is
coherently splitted into two spatial modes $\left\vert a\right\rangle $ and $%
\left\vert b\right\rangle $, remains in the state $\left\vert \varphi
\right\rangle =\alpha \left\vert a\right\rangle +\beta \left\vert
b\right\rangle $ by a polarization independent Mach-Zehnder interferometer.
The phase setting $\left( \alpha ,\beta \right) $ is determined by the
optical path-length difference.

\textit{Step }(C) --- Critical to the third step of our scheme is the
following identity:%
\begin{align}
\left\vert \varphi \right\rangle \left\vert \Psi \right\rangle _{ap}&
=\left( \alpha \left\vert a\right\rangle +\beta \left\vert b\right\rangle
\right) \otimes \left( \left\vert H\right\rangle \left\vert 1\right\rangle
_{a}+\left\vert V\right\rangle \left\vert 0\right\rangle _{a})/\sqrt{2}%
\right)  \notag \\
& =\frac{1}{2}(\left\vert \Psi ^{+}\right\rangle \left\vert \widetilde{%
\varphi }\right\rangle +\left\vert \Psi ^{-}\right\rangle \hat{\sigma}%
_{z}\left\vert \widetilde{\varphi }\right\rangle  \notag \\
& +\left\vert \Phi ^{+}\right\rangle \hat{\sigma}_{x}\left\vert \widetilde{%
\varphi }\right\rangle +\left\vert \Phi ^{-}\right\rangle \left( i\hat{\sigma%
}_{y}\right) \left\vert \widetilde{\varphi }\right\rangle )\text{,}
\label{4}
\end{align}%
where $\left\vert \Psi ^{\pm }\right\rangle =$ $\left( \left\vert
H\right\rangle \left\vert b\right\rangle \pm \left\vert V\right\rangle
\left\vert a\right\rangle \right) /\sqrt{2}$ and $\left\vert \Phi ^{\pm
}\right\rangle =\left( \left\vert H\right\rangle \left\vert a\right\rangle
\pm \left\vert V\right\rangle \left\vert b\right\rangle \right) /\sqrt{2}$
denote the four polarization-spatial Bell states, $\left\vert \widetilde{%
\varphi }\right\rangle =\alpha \left\vert 0\right\rangle _{a}+\beta
\left\vert 1\right\rangle _{a}$. We make a joint Bell state measurement on
the polarization-spatial qubits. To achieve the BSM, the two spatial modes
are combined in PBS1, and the polarization of photon will be analyzed in
each output (see Fig. 1(b)). The half-wave plate (HWP) performs a Hardmard
rotation $\left\vert H\right\rangle \rightarrow \left( \left\vert
H\right\rangle +\left\vert V\right\rangle \right) /\sqrt{2},\left\vert
V\right\rangle \rightarrow \left( \left\vert H\right\rangle -\left\vert
V\right\rangle \right) /\sqrt{2}$ on the polarization modes. If the detector
$D_{1}$ has a click, then it denotes identification of one of the Bell state
$\left\vert \Psi ^{+}\right\rangle =\left( \left\vert H\right\rangle
\left\vert b\right\rangle +\left\vert V\right\rangle \left\vert
a\right\rangle \right) /\sqrt{2}$. Because after the combining of the two
spatial modes on PBS1, the photonic orthogonal polarization can be
coherently superposed into $\left( \left\vert H\right\rangle +\left\vert
V\right\rangle \right) /\sqrt{2}$, which is rotated to $H$ by HWP1 and then
triggers the detector $D_{1}$. Accordingly, the click of the detector $D_{2}$%
, $D_{3}$ and $D_{4}$ corresponds to $\left\vert \Psi ^{-}\right\rangle
,\left\vert \Phi ^{+}\right\rangle $ and $\left\vert \Phi ^{-}\right\rangle $%
, respectively. After the BSM, i.e. the confirmation of only one click from
the single photon detectors, the encoded information is transferred and
encoded into the DFS of atomic ensembles.

\textit{Step }(D) --- According to the outcome of BSM, in the standard
teleportation, a proper local Pauli unitary operation should be carried out
on the atomic quantum memory to recover the original information. However,
it is worth noticing the difficulty of laser manipulation to get the unitary
operation of atomic ensembles. Thanks to the ease of performing precise
unitary transformation on photon, we take the manner of marking instead of
the recovering operation in the standard teleportation technique. The
quantum memory can be made four different marks in a classic way, depending
on the relevant BSM results. When there is a need to retrieve the
information from the quantum memory, we can simultaneously shine a weak
retrieval pulse with suitable frequency and polarization \cite{20} into the
atomic ensembles. The emitted anti-Stokes fields are then combined on PBS.
As a result, the atomic qubit is converted back to single photon qubit. The
efficiency of the transfer is close to unity at a single quantum level owing
to the collective enhancement. Finally, we just apply corresponding unitary
operation in agreement with the prior mark to recover the information. In
fact, the information storing in the atomic ensembles is just the original
one up to the local Pauli operation. The real formation $\left( \alpha
,\beta \right) $ we stored has not been changed. The recovering operation is
just suspended to facilitate the experimental realization within our
technique.

\begin{figure}[tbp]
\includegraphics[width=5cm]{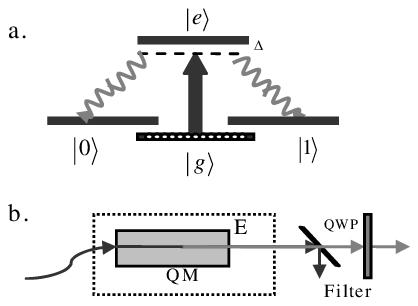}
\caption{(a) The relevant atomic level structure in the ensembles with $%
\left\vert g\right\rangle $ the ground state, $\left\vert e\right\rangle $
the excited state, and $\left\vert 0\right\rangle $ and $\left\vert
1\right\rangle $ the metastable state for storing a qubit of information.
The transition $\left\vert g\right\rangle \rightarrow \left\vert
e\right\rangle $ can be coupled through the classical laser pulse with the
Rabi frequency\textit{\ }$\Omega (t)$ and a detuning $\mathbf{\bigtriangleup
}$, and the forward-scattered Stokes photons come from the transition $%
\left\vert e\right\rangle \rightarrow \left\vert 0\right\rangle $ and $%
\left\vert e\right\rangle \rightarrow \left\vert 1\right\rangle $, which are
left- and right-circularly polarized. (b) The forward-scattered Stokes
photon is collected after the filter which is frequency selective to
separate the pumping light, and transmits the quarter wave plate (QWP). Then
the entanglement of atomic ensemble and photon is generated.}
\end{figure}

On the other hand, the two atomic ensembles $L$ and $R$ can also be replaced
by one ensemble, but with two metastable states $\left\vert 0\right\rangle $
and $\left\vert 1\right\rangle $ to store the quantum information shown in
Fig. 2. The states $\left\vert g\right\rangle $, $\left\vert 0\right\rangle $
and $\left\vert 1\right\rangle $ correspond to the hyperfine or the Zeeman
sublevels of alkali atoms in the ground-state manifold, and $\left\vert
e\right\rangle $ corresponds to an excited state. The $N$ atoms are
initially prepared in the ground state $\left\vert G_{a}\right\rangle $ $%
=\otimes _{i}\left\vert g\right\rangle _{i}$. The transition $\left\vert
g\right\rangle \rightarrow \left\vert e\right\rangle $ is driven
adiabatically by a weak classical laser pulse with the corresponding Rabi
frequency denoted by $\Omega (t)$. With the short off-resonant driving
pulse, only one atom is transferred nearly with unit probability to the
excited state $\left\vert e\right\rangle $. The excited state will transit
into the metastable states $\left\vert 0\right\rangle $ or $\left\vert
1\right\rangle $ with equal probabilities by emitting a left- or
right-circularly polarized Stokes photon in the forward direction. Such
emitting events are uniquely correlated with the excitation of the symmetric
collective atomic mode $S_{h}$ which is given by $S_{h}=$ $\left( 1/\sqrt{%
N_{a}}\right) \sum_{i}\left\vert g\right\rangle _{i}\left\langle
h\right\vert $ $(h=0,1)$. The emission of single Stokes photon will result
in the state of atomic ensembles by $\left\vert h\right\rangle
_{a}=S_{h}^{+}\left\vert 0_{a}\right\rangle .$We also can define single mode
bosonic operator $a_{h}$ for the Stokes pulse with its vacuum state denoted
by $\left\vert 0\right\rangle _{p}$. The emitting process can be defined by $%
\left\vert L\right\rangle =a_{0}^{+}\left\vert 0_{p}\right\rangle $ and $%
\left\vert R\right\rangle =a_{1}^{+}\left\vert 0_{p}\right\rangle $, $%
\left\vert L\right\rangle $ and $\left\vert R\right\rangle $ denote the
polarization of single Stokes photon. Before the emitted Stokes photon is
coupled into the fiber, it transmits a quarter wave plate (QWP). By
neglecting the small vacuum component and the high order terms, the
entangled state of the composite of atomic ensemble and photon can be
written into the form

\begin{equation}
\left\vert \Psi \right\rangle _{ap}=\left(
P_{V}^{+}S_{0}^{+}a_{0}^{+}+P_{H}^{+}S_{1}^{+}a_{1}^{+}\right) /\sqrt{2}%
\left\vert vac\right\rangle _{ap},  \label{5}
\end{equation}%
with $\left\vert vac\right\rangle _{ap}=\left\vert G_{a}\right\rangle
\left\vert 0_{p}\right\rangle $, i.e. $\left\vert \Psi \right\rangle
_{ap}=\left( \left\vert H\right\rangle \left\vert 1\right\rangle
_{a}+\left\vert V\right\rangle \left\vert 0\right\rangle _{a}\right) \sqrt{2}
$. Via the above teleportation, the information will be stored into the
atomic memory. Due to the collective enhanced coherent interaction, the
atomic metastable states $\left\vert 0\right\rangle _{a}$ and $\left\vert
1\right\rangle _{a}$ can be transferred to optical excitations with high
efficiency. So the information can be read out for further quantum
information processing.

\begin{figure}[tbp]
\includegraphics[width=8cm]{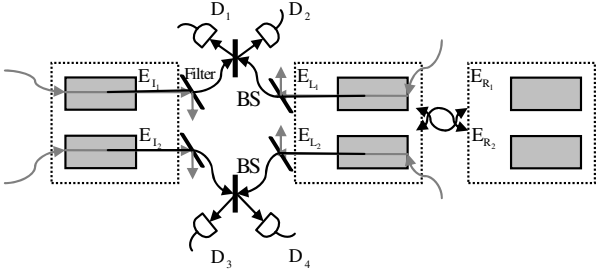}
\caption{Schematic setup for remote transfer of unknown quantum state. The
unknown quantum state $(\protect\alpha S_{I_{2}}^{+}+\protect\beta %
S_{I_{1}}^{+})\left\vert 0_{a}0_{a}\right\rangle _{I_{1}I_{2}}$ is prepared
in the atomic ensemble pair $I$. The atomic ensemble pair $L$ and $R$ are in
a long-distance entangled state $\left\vert \Psi \right\rangle _{LR}$. After
shined simultaneously by the repump pulse, the collective atomic excitations
in the atomic ensemble $I$ and $L$ are transferred to the optical
excitations, which are registered by the sigle-photon detectors.}
\end{figure}

After the atomic quantum memory has been established, it is very necessary
to consider its use in quantum communication protocols. Take remote transfer
of unknown quantum state as an example. Suppose that a long-distance
entangled state $\left\vert \Psi \right\rangle _{LR}$ between the double
pairs of atomic ensembles $L_{1}$, $L_{2}$ and $R_{1}$, $R_{2}$ is generated
by the quantum repeater \cite{21,22}, $\left\vert \Psi \right\rangle
_{LR}=(S_{L_{1}}^{+}S_{R_{2}}^{+}+S_{L_{2}}^{+}S_{R_{1}}^{+})/\sqrt{2}%
\left\vert vac\right\rangle _{LR}$, where $\left\vert vac\right\rangle
_{LR}=\left\vert 0_{a}0_{a}0_{a}0_{a}\right\rangle _{L_{1}L_{2}R_{1}R_{2}}$.
We denote the entangled logical state by $\left\vert \Psi \right\rangle
_{LR}=\left( \left\vert 0\right\rangle _{L}\left\vert 1\right\rangle
_{R}+\left\vert 1\right\rangle _{L}\left\vert 0\right\rangle _{R}\right) /%
\sqrt{2}$, where $\left\vert 0\right\rangle _{L(R)}=\left\vert
0\right\rangle _{L_{1}(R_{1})}\left\vert 1\right\rangle _{L_{2}(R_{2})}$, $%
\left\vert 1\right\rangle _{L(R)}=\left\vert 1\right\rangle
_{L_{1}(R_{1})}\left\vert 0\right\rangle _{L_{2}(R_{2})}$. The unknown
quantum state storing in the two atomic ensembles $I_{1}$ and $I_{2}$ is $%
\left\vert \widetilde{\varphi }\right\rangle _{I}=(\alpha
S_{I_{2}}^{+}+\beta S_{I_{1}}^{+})\left\vert 0_{a}0_{a}\right\rangle
_{I_{1}I_{2}}=\alpha \left\vert 0\right\rangle _{I}+\beta \left\vert
1\right\rangle _{I}$. Then shined simultaneously by the repump pulse which
is near-resonant with the atomic transition $\left\vert s\right\rangle
\rightarrow \left\vert e\right\rangle $, the collective atomic excitations
in the ensembles $I_{1}$, $L_{1}$ and $I_{2}$, $L_{2}$ are transferred to
the optical excitations, which, interfere respectively in a 50\%-50\% beam
splitter, and are detected by two single-photon detectors on each output. If
one click in $D_{1}$ or $D_{2}$, and one click in $D_{3}$ or $D_{4}$ have
been confirmed, our protocol succeeds in transferring the unknown quantum
state in the ensemble pair $I$ into the ensemble pair $R$ up to a local $\pi
$-phase rotation. The fidelity of the remote transfer protocol is nearly
perfect.

\begin{figure}[tbp]
\includegraphics[width=8cm]{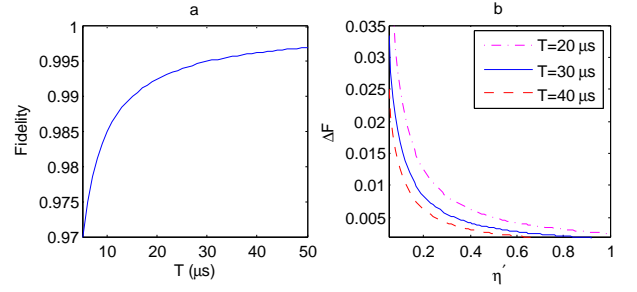}
\caption{(color online) \ (a) The improvement of quantum memory fidelity
with the average preparation time. Here we choose the overall efficiency of
photon detector, the optical appratus and fiber $\protect\eta ^{\prime }=1/3$%
. (b) The fidelity imperfection versus the overall efficiency $\protect\eta %
^{\prime }$ with the average preparation time $T=20$ $\protect\mu s,$ $30$ $%
\protect\mu s,$ $40$ $\protect\mu s$. }
\end{figure}

Finally, we briefly discuss the influence of practical noise and
imperfections on the memory. The fidelity of the information we store in the
atomic ensembles is mainly depend on the purity of entangled state. In
practice, the entanglement between the atomic ensembles and the Stokes
photon is in fact a mixed entangled state. The dominant noise arises from
the channel attenuation, spontaneous photon scattering into random
directions, coupling inefficiency for the channel, and inefficiency of
single-photon detectors. Taking into account these noises, as will be shown
below, the entangled state can be modified to

\begin{equation}
\rho _{ap}=p_{0}\left\vert vac\right\rangle _{ap}\left\langle vac\right\vert
+p_{1}\left\vert \Psi \right\rangle _{ap}\left\langle \Psi \right\vert
+p_{o}\rho _{o}\text{,}  \label{6}
\end{equation}%
where $p_{0}$, $p_{1}$ and $p_{o}$ are the probabilities of the vacuum state
that the two atomic ensembles are in the ground state, one-excited state and
the states with more excitations. Then we will give the probability an
analysis. $\left( i\right) $ We assume the probability of creating a Stokes
photon behind PBS is $2p_{c}$, the detection efficiency $\eta $ \cite{23} is
determined by the finite photon-collection (coupling) efficiency of the
optical apparatus, $\chi $, and by the quantum efficiency of the detector
itself, i.e. the efficiency of distinguishing single photon click event from
more photons, $\eta _{d}$, so the success probability of getting a
single-photon detector click is determined as $p_{1}\approx 2p_{c}\chi \eta
_{d}e^{-L_{0}/L_{att}}$, where we have considered the channel attenuation
factor as $e^{-L_{0}/L_{att}}$ \cite{15}, $L_{att}$ is the channel
attenuation length, and the other noise is independent of the communication
length $L_{0}$. $\left( ii\right) $ In the entanglement generation step, we
have neglected the influence of the detector dark counts which is denoted by
$p_{dc}$ in each Raman round. In fact, it contributes to the vacuum
component with the probability $p_{0}\approx p_{dc}/(p_{c}\eta ^{\prime }),$
$\eta ^{\prime }=\eta e^{-L_{0}/L_{att}}$ denotes the overall efficiency of
the single-photon detector, the optical apparatus and channel in the scheme.
Now, we take it into account. However, this vacuum component is typically
very small since the normal dark count rate (100 Hz) is much smaller than
the repetition frequency (10 MHz) of the Raman pulses \cite{15}. In
addition, finally this component will be automatically eliminated in our
scheme since its effect can be included by the detector inefficiency in the
application measurements. $\left( iii\right) $ If more than one atom is
excited to the collective mode S, due to the inefficiency of the photon
detector, there is only one click. The probability of this event is given by
$p_{o}\sim p_{c}^{n}\chi (1-\eta _{d})e^{-L_{0}/L_{att}}$ (decays
exponentially with the number of excited atoms $n$). So, the fidelity
between the generated state $(6)$ and the ideal state $(3)$ is decreased by
the high-order component which is proportional to $p_{c}$, we can simply
estimate the fidelity imperfection of our quantum memory $\Delta
F=1-F\approx p_{c}$. Decreasing the small controllable excitation
probability $p_{c}$ for each driving Raman pulse, we can make the fidelity
of the information close to one, but the longer preparation time $T$ is
cost. We have to repeat the process about $1/p_{1}$ times, with the total
average preparation time $T\sim 1/\left( p_{1}f_{p}\right) $, where $f_{p}$
is the repetition frequency of the Raman pulses. As evident from Fig. 4(a),
the fidelity of quantum memory can exceed $0.99$ for the average preparation
time $T>15$ $\mu s$. Furthermore, the fidelity is insensitive to the
variation of the overall efficiency $\eta ^{\prime }$ caused by the
single-photon detector, the optical apparatus and channel shown in Fig.
4(b). For instance, the change of the fidelity is about $10^{-2}$ for $\eta
^{\prime }$ varying from $0.1$ to $1$. The Fig. 4(b) also shows that the
fidelity is more insensitive to the overall efficiency $\eta ^{\prime }$
with the increase of the average preparation time.

In summary, resorting to the idea of quantum teleportation, we have proposed
a quantum memory scheme to imprint the quantum state of the photon into the
DFS of atomic ensembles with high fidelity, and show one of applications
with the memory. Moreover, based on the current technology of laser
manipulation, photonic state operation through optical elements, and
single-photon detection with moderate efficiency, the scheme is inherently
resilient to the noise. Due to the long coherent time of atomic ensemble
\cite{13,14}, the teleported information can finally be read out for further
quantum information applications.

\bigskip This work was supported by the National Natural Science Foundation
of China under Grant. Nos. 10404007 and 60578055, and the State Key
Development Program for Basic Research of China (Grant No. 2007CB925204 and
2009CB929604). \ \ \ \ \ \ \ \ \ \ \ \ \ \ \ \ \ \ \ \ \ \ \ \ \ \ \ \ \ \ \
\ \ \ \ \ \ \ \ \ \ \ \ \ \ \ \ \ \ \ \ \ \ \ \ \ \ \ \ \ \ \ \ \ \ \ \ \ \
\ \ \ \ \ \ \ \ \ \ \ \ \ \ \ \ \ \ \ \ \ \ \ \ \ \ \ \ \ \ \ \ \ \ \ \ \ \
\ \ \ \ \ \ \ \ \ \ \ \ \ \ \ \ \ \ \ \ \ \ \ \ \ \ \ \ \ \ \ \ \ \ \ \ \ \
\ \ \ \ \ \ \ \ \ \ \

\ \ \ \ \ \ \ \ \ \ \ \ \ \ \ \ \ \ \ \ \ \ \ \ \ \ \ \ \ \ \ \ \ \ \ \ \ \
\ \ \ \ \ \ \ \ \ \ \ \

\bigskip\

\bigskip


\begin{thebibliography}{99}
\bibitem{17} Here, we prepare the etanglestate by the application
measurement. If in the final measurement, we keep only the results for which
an excitation appears on one of the esembles $L$ and $R$, i.e. postselect
the case when only one detector registers a click, the whole of our scheme
is successful since only the state $(2)$ has contributions to the
measurement.
\end{thebibliography}
\end{document}